# A Theoretical Study on Crystallization of Polyethylene Using Modified Weighted Density Approximation (MWDA)

A. Razeghizadeh[1,*], V. Rafee[1]
[1]Department of Physics, Faculty of science, University Payame Noor, Iran
Email of the corresponding author: razeghizadeh@yahoo.com

**Abstract**
In this article, the crystallization of Polyethylene is investigated by the modified weighted density approximation. Also a direct correlation function of Polyethylene based on the RISM theory is used. The free energy of a Polyethylene is calculated using density functional theory. The crystallization together with solid and liquid densities are calculated and finally compared with the PRISM simulations and experimental results. The result obtained by modified weighted density approximation MWDA is shown to be in a better agreement with the experimental results rather than the PRISM simulations.

**1- Introduction**
A polymer is a very large molecule that is made by repetitions of a unit. The units are linked together in such a way that the polymer structure can be made. The crystallization ability of some of this material has a very important industrial application [1].
Most of the factors that affect the crystallization rate of polymers are polymer chain structure, molecular weight, the ability to form a secondary valence band and the thermal history of the polymer during construction.
A large number of these factors made calculations very difficult and made a variety of the studying methods.
Under certain conditions, most of these polymers will crystallize into FCC and find special properties. Many efforts have been taken place to explain this phenomenon [2].
Many different analytical methods have been developed to describe the structural properties and phase transition of many-body particle systems. These methods included different models such as Flory lattice [3-5], growth theories [6, 7], Monte Carlo simulation [8-10], Landau-de Gennes theory [11] and density functional theory [12-16]. This article indicates that liquid simple model can be used for polymers.
The density functional theory in the form of classic [12] and quantum [13] is used to study different thermodynamic properties of condensed matter for simple liquid.
This theory is used to study homogeneous systems [14-16] and inhomogeneous such as limited liquid, liquid-solid interface [17, 18], crystals [19-21], electron cloud distribution in atoms [22], phonon dispersions of cluster crystals [23], calculations of the electronic structure in atoms [24], the electronic structure of solids [25], phase transition [26], too.
The basic assumption to study the phase transition in all of density functional theories is that the thermodynamic potential of a non-uniform system is approximated to its equivalent uniform system.
The detail approximations used in relationships distinguish these theories. Therefore, different approaches have been used earlier. The approximations Rama Krishnan and Yussouff (RY), is one of the approximations that can be mentioned [27]. This approximation is suitable to study, freezing transition simple and mixed liquid [28, 29]. Another approximation in this field is a weighted density approximation (WDA) [30-34]. This approximation will lead to a better result compared to RY approximation [35-38]. Unfortunately, since calculations in this method are very difficult and very long calculation as using this is very difficult. Therefore, we present modified weighted density approximation (MWDA) [39, 40] that first this method has suitable resulted and second, since this method uses a classical approximation, calculation in method is easy [41]. Unfortunately the large amount of calculations in this method makes it usage difficult.

Finally modified weighted density approximation (MWDA) [39, 40] is mentioned because this method has very good accuracy and since this method uses a classical approximation therefore it has a suitable amount of calculations [41].

Some efforts were made to use density functional theory to explain polymers that a few of them can be mentioned.

Yethiraj and Woodward presented a density functional for polymer based on WDA theory [42]. A few years later, Yethiraj corrected his own theory by using complex weighted functional [43]. He studies in his own article two models: A freely jointed hard chain model and a fused-sphere hard chain model. He used correlation function obtained from the polymer reference interaction site model (PRISM) and Ornstein-Zernike equations for polymer to start own calculation.

Different studies have been done on polymers, especially polyethylene, by density functional method [44, 45]. As an example, McCoy et al. studied polymer system by density functional [46].

In this paper, at first, we calculate the correlation function of polyethylene ($N = 6429$) by curve fitting of experimental correlation function in [46]. Then we calculate different parameters of polyethylene such as solid and liquid phase density, Lindemann criterion, chemical potential, monomers diameter and isothermal compressibility of the liquid phase of polyethylene. After that we compared our results with experimental ones and the results of different methods. Finally, we show that our results of MWDA method have a better agreement with experimental results in comparison with more complex methods such as Monte Carlo simulation or WDA, etc. Therefore, for the first time, we show that different parameters of polyethylene by MWDA method can calculate using correlation function, simply.

## 2- Theoretical formalism
### 2-1- Weighted density functional formalism

The basis of density functional theory is determining a grand free energy $\Omega[\rho]$ as an individual function of the density of single-particle $\rho(\vec{r})$. The goal of this part is to investigate thermodynamic properties of homogeneous systems.

At the first, The grand free energy $\Omega[\rho]$ is obtained as [43]:

$$\Omega[\rho] = F[\rho(\vec{r})] + \int d\vec{r}\rho(\vec{r})[V_{ext}(\vec{r}) - \mu] \qquad (1)$$

Where $\mu$ is chemical potential and $V_{ext}(\vec{r})$ is the external potential and $F[\rho(\vec{r})]$ is single particle Helmholtz free energy functions of classical many-body system.

The equilibrium density of the system is calculated by minimizing $\Omega[\rho]$ with respect to $\rho(\vec{r})$ in DFT method. The macroscopic properties of the system are calculated in an equilibrium state [43].

$$\frac{\delta \Omega[\rho]}{\delta \rho(\vec{r})} = 0 \qquad (2)$$

The equilibrium Helmholtz free energy of the system can be calculated as [12]:

$$F[\rho_0] = F_{int}[\rho_0] + \int \rho_0(\vec{r}) V_{ext}(\vec{r}) d\vec{r} \qquad (3)$$

Where $F_{int}[\rho_0]$ is intrinsic Helmholtz free energy of the system.
Therefore, from Eq. (2) we have [12]:

$$\mu = \mu_{int}[\rho_0, \vec{r}] + V_{ext} \qquad (4)$$

Where $\mu_{int}$ is the intrinsic chemical potential and it is calculated as [12]:

$$\mu_{int} = \left. \frac{\delta F[\rho]}{\delta(\rho)} \right|_{\rho=\rho_\circ}$$

Also the Helmholtz free energy of a classical many-body system is obtained as [43]

$$F[\rho(\vec{r})] = F_{id}[\rho(\vec{r})] + F_{ex}[\rho(\vec{r})] \qquad (5)$$

Where $F_{id}[\rho(\vec{r})]$ is ideal part (It is the contribution of ideal gas) and $F_{ex}[\rho(\vec{r})]$ is excess part (It is contribution of molecular interactions).

It is a standard method to calculate Helmholtz free energy $F[\rho]$ that one separates ideal part $F_{id}[\rho(\vec{r})]$ and make some approximations for the excess part $F_{ex}[\rho(\vec{r})]$.

The ideal part is calculated as [43]:

$$F_{id}[\rho(\vec{r})] = \beta^{-1} \int d\vec{r} \rho(\vec{r})[\ln \rho(\vec{r}) - 1] + \int d\vec{r} V_{int}(\vec{r}) \rho(\vec{r}) \qquad (6)$$

Where $V_{int}(\vec{r})$ is intramolecular interaction. The second term in Eq. (6) can be neglected because in MWDA and WDA long-range effect is negligible.

The excess Helmholtz free energy $F_{ex}[\rho]$ of a classical many-body system is a unique function of density and it is shown below [36]:

$$F_{ex}[\rho] = \int d\vec{r} \rho(\vec{r}) \phi_{ex}(\vec{r};[\rho]) \qquad (7)$$

Where $\phi_{ex}(\vec{r};[\rho])$ is the excess free energy per particle. In weighted density functional approximation, $F_{ex}[\rho]$ is obtained as [36]:

$$F_{ex}^{WDA}[\rho] = \int d\vec{r} \rho(\vec{r}) \varphi_0 \overline{\rho}(\vec{r}) \qquad (8)$$

Where $\varphi_0$ is the excess free energy per particle of homogeneous liquid.
$\overline{\rho}(\vec{r})$ is the mean weighted density of per particle $\rho(\vec{r})$ regarding weighted functional $\omega$ and it is calculated like this [36]:

$$\overline{\rho}(\vec{r}) = \int d\vec{r}' \rho(\vec{r}') \omega[\vec{r} - \vec{r}'; \overline{\rho}(\vec{r})] \qquad (9)$$

Weighted functional satisfies a normalization condition:

$$\int d\vec{r}' \omega[\vec{r} - \vec{r}'; \overline{\rho}(\vec{r})] = 1 \qquad (10)$$

Also the relationship between correlations function $N$ particles and $F_{ex}[\rho]$ is defined as below [36, 47]:

$$C^{(n)}(\vec{r}_1, ..., \vec{r}_n; [\rho]) = -\frac{\delta^n \beta F_{ex}[\rho]}{\delta \rho(\vec{r}_1)...\delta \rho(\vec{r}_n)} \qquad (11)$$

## 2-2- Modified Weighted Density Approximation

In this article the modified weighted density functional approximation is used because it is simple and is in a better agreement with simulations [47]. The weighted density approximation method is approximated for $\phi_{ex}(\vec{r},[\rho])$. This method makes it possible to calculate the localized excess free energy per particle.

Also formulation focuses on excess free energy per particle $\beta F_{ex}/N$; in modified weighted density approximation method which $N$ is the number of particles of the system.

Since $F_{ex}[\rho]/N$ does not depend on the position of the particle compared with $\phi_{ex}(\vec{r},[\rho])$, we show the weighted density by $\hat{\rho}$ that difference between it and $\overline{\rho(\vec{r})}$ is independence position.

The approximation used in calculation of $F_{ex}[\rho]/N$ can be written as [36]:

$$F_{ex}^{MWDA}[\rho]/N = \varphi_{ex}(\hat{\rho}) \tag{12}$$

Where:

$$\hat{\rho} \equiv \frac{1}{N}\int d\vec{r}\rho(\vec{r})\int d\vec{r}'\rho(\vec{r}')\tilde{\omega}(\vec{r}-\vec{r}';\hat{\rho}) \tag{13}$$

The weight density functional satisfies the normalization condition in Fourier space [48]:

$$\tilde{\omega}(k;\rho_0) = \frac{-1}{2\varphi'(\rho_0)}[\beta^{-1}C^{(2)}(k;\rho_0) + \delta_{k,0}\rho_0\varphi_0''(\rho_0)] \tag{14}$$

The weighted density $\hat{\rho}$ must be calculated for study of the Helmholtz free energy. This calculation is simple in Fourier space, because in Eq. (13) the volume integrals convert to summation on inverse lattice vectors [48]:

$$\hat{\rho} = \frac{1}{N}\int d\vec{r}\rho(\vec{r})\int d\vec{r}'\rho(\vec{r})\,\omega(\vec{r}-\vec{r}';\hat{\rho}) = \rho_s + \frac{1}{\rho_s}\sum_{G\neq 0}\rho_G^2\,\omega_G(\hat{\rho}) \tag{15}$$

Where $\rho_s$ is mean solid density, $\rho_G$ and $\tilde{\omega}_G$ are a Fourier component of solid density and weighting function respectively and Fourier component $\rho_G = \rho_s e^{-G^2/4\alpha}$.

We can be assumed $\rho_s(\vec{r})$ focused as a summation of normalization of Gaussian on the $\vec{R}$ situation of lattice in FCC lattice.

Therefore $\rho_s(\vec{r})$ can be written as [36]:

$$\rho_s(\vec{r}) = (\frac{\alpha}{\pi})^{3/2}\sum_R \exp[-\alpha(\vec{r}-\vec{R})^2] \tag{16}$$

Where $\alpha$ is localized parameter and it is $\alpha = 0$ for homogeneous liquid limit and atoms is localized in liquid-to-solid phase transition with $\alpha$ increase.

$\hat{\rho}$ can be written as a below with replacing $\rho_G, \tilde{\omega}_G$ in Eq. (15) [48]:

$$\hat{\rho}(\rho_s,\alpha) = \rho_s[1-\frac{1}{2\beta\varphi_0'(\rho)}\sum_{G\neq 0}\exp(\frac{-G^2}{2\alpha})C^{(2)}(\vec{G};\hat{\rho}) \tag{17}$$

The Helmholtz free energy per particle and direct correlation function is needed for calculation of $\hat{\rho}$.

In this article we used a direct correlation function of polyethylene ($N = 6429$) [48, 49].

$\varphi_0$ defined as below with PY approximation [48, 49]:

$$\varphi_0(\eta) = (3/2)[(1/(1-\eta)^2)-1]-\ln(1-\eta) \qquad (18)$$

Where $\eta = \pi\rho\sigma^3/6$

For calculation of ideal part, $F_{id}[\rho]/N$ can be written as [48]:

$$F_{id}[\rho_s] = \beta^{-1}\int d\vec{r}\rho_s(\vec{r})\{\ln[\rho_s(\vec{r})\lambda^3]-1\} \qquad (19)$$

For $\alpha > 50$ it can be using fallow [48]:

$$\frac{\beta F_{id}(\rho_s,\alpha)}{N} = \frac{3}{2}\ln\left(\frac{\alpha}{\pi}\right)+3\ln(\lambda)-\frac{5}{2} \qquad (20)$$

Where $\lambda = 0.006/\rho_s$ is heat wavelength of the system.

Helmholtz free energy per particle of classical many-body system calculated as below with the calculation of $F_{id}[\rho]/N, F_{ex}[\rho]/N$ [48, 49]:

$$\frac{\beta F}{N} = \frac{\beta F_{id}}{N}+\frac{\beta F_{ex}}{N} \qquad (21)$$

In the liquid phase, we used the Carnahan-Starling approximation for calculation, because it is exactly against PY approximation for the liquid phase, therefor we can written [50]:

$$\varphi_\circ^{cs}(\eta) = \frac{2}{(1-\eta)}+\frac{1}{(1-\eta)^2}-3 \qquad (22)$$

$\beta F_{ex}/N$ is calculated by Eq. (14), Eq. (22) for liquid phase. $F_{id}$ is calculated by replacing liquid density to solid density in (19) for liquid phase and integrated, that become [36]:

$$\frac{F_{id}[\rho_s]}{V} = \beta^{-1}\rho_s\{\ln[\rho_s]-1\} \qquad (23)$$

**3- Results and discussions**
**3-1- Direct correlation function of polyethylene**
The direct correlation function must be replaced with the system's potential in the density function theory formalism. The direct correlation function for hard spheres Potential [30-36], Lennard-Jones potential [51] etc. has been calculated earlier.
In this paper, the direct correlation function calculates with using the Fig. (1) for polyethylene [46].
We calculate the direct correlation function $C(r)$ and Fourier transform it $C(k)$ by experimental Curve Fitting polyethylene $(N = 6429)$ [46].
Now, the direct correlation function as a polynomial of degree 3 is obtained by drawing the function, fitted with an error of less than 0.03 which show in Fig. (1).

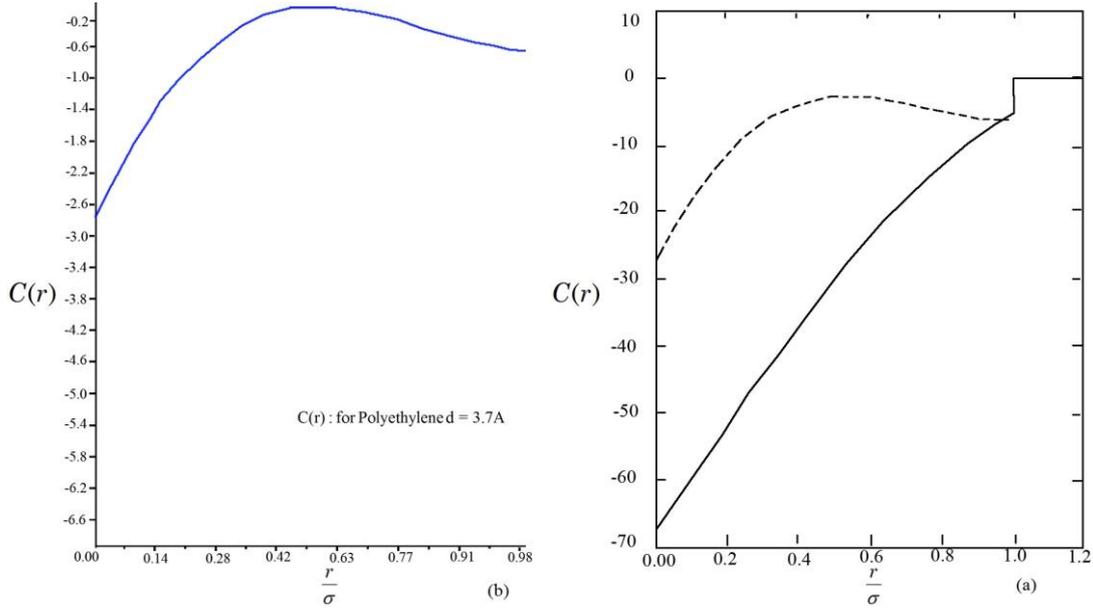

**Fig 1.** (a) Direct correlation function for polyethylene (dashed line; polymer RISM theory; $T = 413°K$ and hard spheres solid line; Percus-Yevick theory) at the phase transition [46]. (b) Fitness of Direct correlation function of polyethylene

First, we write the correlation function as a polynomial of degree 3:
$$C(r) = a + b(r/\sigma) + c(r/\sigma)^2 + d(r/\sigma)^3 \qquad (24)$$
Where $a = -2.70000$, $b = 47.27416$, $c = -278.34645$ and $d = 484.09578$. then by using Fourier transformation and replacing $\rho$ by $\eta$, correlation function is obtained Fourier transform as:
$$C(k) = 48[-b/k^4 + 12d/k^6]\eta \qquad (25)$$
Fig. (2) shows Fourier transforms of the correlation function.

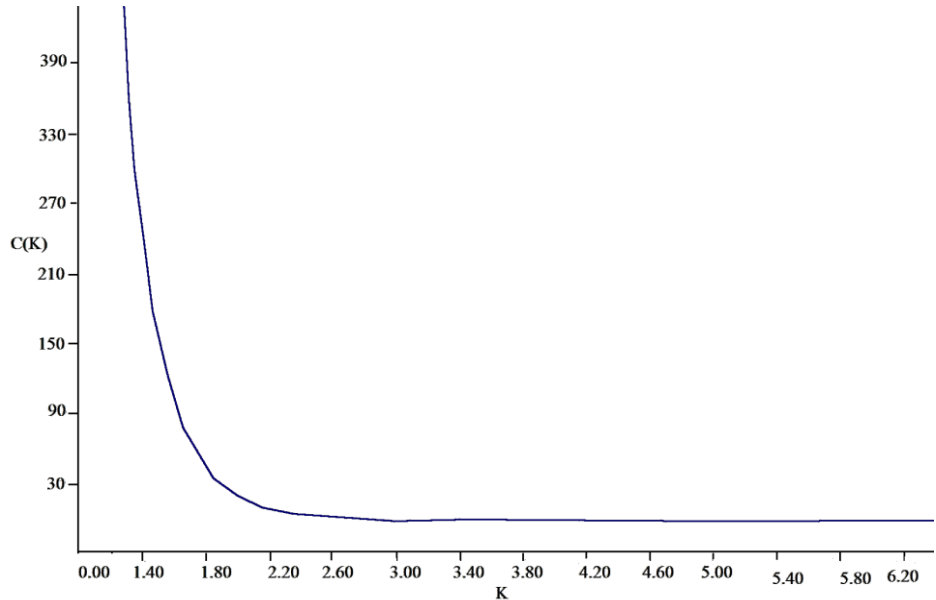

**Fig 2.** Plot C(K) for crystallization of polyethylene ($N$=6429, $\rho_s = 0.83$)

### 3-2- Monomers diameter polyethylene

The researcher studied different lot of $C(r), C(k)$ and founded a better agreement for MWDA method is for hard sphere diameter $\sigma_{CH_2} = 3.7 A°$ that this volume, according to the experimental result $\sigma_{CH_2}$ [52] of Chandler result and then Slonimskii result [53] that $\sigma = 4.92 A°$, is better.

The exact amount of the monomer calculated by X-Ray and $\sigma = 3.90 A°$ that it has calculated with McCoy [46].

The results are presented in Tab. (1) and in Fig. (1) we compare own results whit McCoy result of the P-RISM method [46].

Table 1. The results of Monomers diameter polyethylene of different methods

| Method | $\sigma$ |
|---|---|
| Chandler [52] | $3.7 A°$ |
| Slonimskii [53] | $4.92 A°$ |
| X-Ray diffraction [46] | $3.90 A°$ |
| This work (MWDA) | $3.7 A°$ |

### 3-3- Liquid and solid phase density of polyethylene

At the first, we selected $\rho_s$ equal 0.9 and we changed it to 1.1 in 0.001unit interval. (Liquid freezing point ranges hard sphere) [48, 49]. After that we calculated lattice constant FCC $a = (4/\rho_s)^{1/3}$ as functions of $\rho_s$. After that we changed $\alpha$ from 1 to 250 as in one unit interval.

Then, inverse lattice vector will be calculated for 6200 consecutive shells using the calculated lattice constant in which the smallest vector $\vec{G}(1) = \frac{2\pi}{a}(1,1,1)$ equal $|\vec{G}(1)| = \frac{2\pi}{a}\sqrt{3}$. Then selected initially $\hat{\rho}$ and calculated $\eta = \pi\hat{\rho}/6$. After that we calculate $C_\circ^{(2)}(\vec{G};\eta)$ for each $\vec{G}$ from 1 to 6200 [54, 55].

$\varphi_\circ'(\hat{\rho})$ will be given from differentiation of $\varphi_\circ(\eta)$ in Eq. (18) with respect to $\hat{\rho}$ and put in that $\eta$. Then $\hat{\rho}(\rho_s,\alpha)$ is calculated with $\varphi_\circ'(\hat{\rho})$ in Eq. (17). Now, we calculate Eq. (17) as self constantly to give $\hat{\rho}$ with requested accuracy. Then we are given $F_{id}[\rho]/N$ by Eq. (20) and $F_{ex}[\rho]/N$ by Eq. (12). Finally, we calculated $\beta F/N$ by Eq. (21).

The Fig. (3) shows $\beta F/N$ as functional of $\alpha$ half the width of the Gaussian. The minimize point in this show equilibrium state of the system.

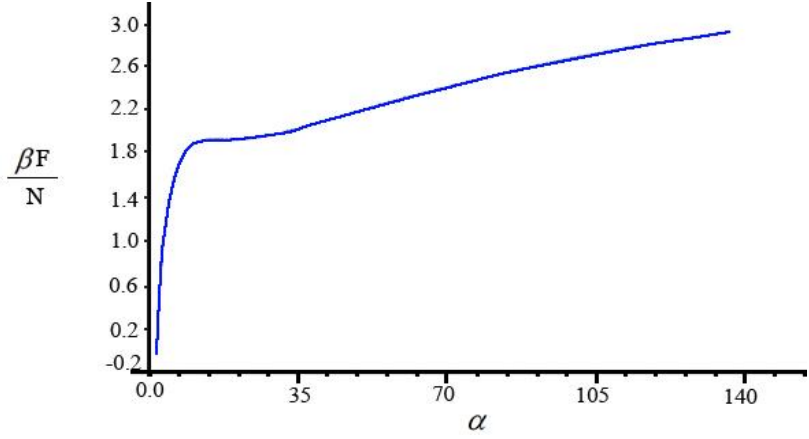

**Fig 3.** The free energy of crystallization phase of polyethylene as a function of $\alpha$ ( $\rho_s = 0.83$ )

Now, for liquid phase, we calculate $\beta F/N$ by calculation of $\varphi_0$ in Eq. (22) and we achieve $\beta F_{ex}/N$. Then $\beta F_{id}/N$ is given by Eq. (23). Finally, we calculate Helmholtz free energy per particle for the liquid phase of Eq. (21).

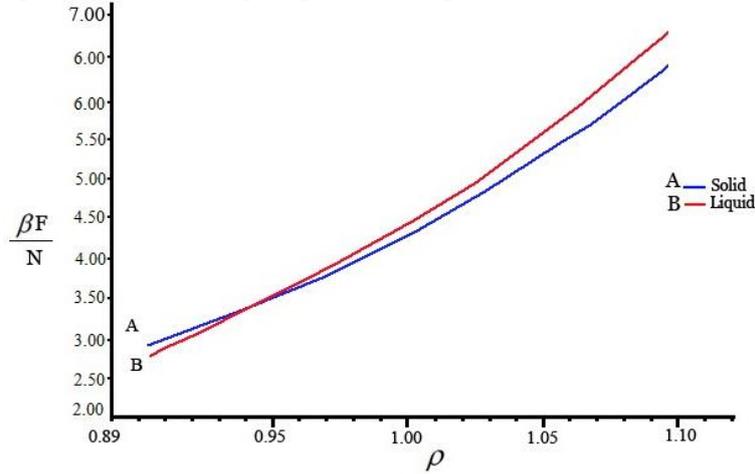

**Fig. 4** The free energy of solid and liquid phase of polyethylene ( $N = 6429$ )
at the point of transition (coexistence)

Fig. (4) shows free energy of solid and liquid phase of polyethylene ( $N = 6429$ ) at the point of transition.

The calculation of solid and liquid phase density is done with two methods. In the first we find tangent to the two curves, the contact points between this tangent and two curves can be counted as a solid and liquid phase in freezing point respectively.

In the second method we get the polynomial of degree 2 for free energy of solid and liquid phase curves. Hence we, using these relations' differentiations, calculate the chemical potential and then thermodynamic potential, according to the following equations [48, 49]:

$$\frac{\mu(P)}{V} = \frac{\partial}{\partial \rho}\left[\frac{F(P)}{V}\right] \quad , \qquad (26)$$

$$w(P) = \frac{F(P)}{V} - P\mu(P) \qquad (27)$$

Then, we equal Eq. (27) with Eq. (28) and it is coexistence condition. Therefore, we will have [36]:

$$\frac{\mu(P_s)}{V} = \frac{\mu(P_l)}{V}$$
$$w(P_s) = w(P_l) \tag{28}$$

$\rho_s$ and $\rho_l$ is obtained by Eq. (28)

**Table. 2** Comparison of solid and liquid phase density of polyethylene in the transition point of different methods.

| Method | $\rho_l$ | $\rho_S$ |
|---|---|---|
| Experimental [56-58] | 0.78 | 0.95 |
| P-RISM [46] | 0.77 | 1.13 |
| MWDA (This Work) | 0.78 | 0.82 |

### 3-4- Isothermal compressibility coefficient of liquid phase of polyethylene

We calculated he isothermal compressibility coefficient of the liquid phase of polyethylene in coexisting condition with crystallization phase inputting related diameter of hard sphere and thermodynamic pressure in isothermal compressibility coefficient relation and differentiating respect to density as follows [59]:

$$\beta P / \rho = (1 + \eta + \eta^2 - \eta^3)/(1-\eta)^3 \tag{29}$$

For calculation compressibility factor polyethylene we inputting $\eta$ in Eq. (29) and using the definition of Eq. (29) and by compressibility coefficient:

$$1/K_T = \rho(\partial p / \partial \rho) \tag{30}$$

Finally, we have: $\beta / K_T = 0.77$

Tab. (3) shows our result compared with McCoy result.

**Table. 3** Comparison of results of compressibility factor polyethylene in liquid phase in different methods.

| Method | $\beta / K_T$ |
|---|---|
| McCoy [46] | 0.70 |
| MWDA (This Work) | 0.77 |

As shown in Tab. (3), the difference between our result and McCoy's is about 10%, which confirms the compatibility mapping to study the linear polymers.

### 3-5- Lindemann criterion and chemical potential and solid and liquid phase packing fraction of polyethylene

The chemical potential of crystal of polyethylene ($N = 6429$) is obtained as bellow with having of packing fraction and diameter monomers $\sigma = 3/7 A°$ [36]:

$$\mu_s[\rho_0, r] = \left.\frac{\delta F[\rho]}{\delta \rho(r)}\right|_{\rho=\rho_0} = 6.083$$

Lindemann criterion is calculated as [36]:

$$L = \sqrt{\frac{3}{\alpha a^2}} = 0.154 \tag{31}$$

The rate of change of Lindemann criterion is a criterion of liquid-to-solid phase transition.

The chemical potential and solid and liquid phase packing fraction calculated by $\pi\rho_s/6, \pi\rho_l/6$ respectively [36].

These quantities are very important for this reason we calculate them in this article.

Unfortunately, these quantities for of polyethylene ($N=6429$) do not calculation earlier. Consequently we cannot compare our results of these quantities by other works (Tab. 4). But because Lindemann criterion and chemical potential and solid and liquid phase packing fraction of polyethylene is as functional of density and so we show the accuracy of density in Tab. (2), we confirm these results.

**Table. 4** The result Lindemann criterion, chemical potential and solid and liquid phase packing fraction of polyethylene by MWDA method

| $\mu_S$ | $\pi\rho_s/6$ | $\pi\rho_l/6$ | $L$ |
|---|---|---|---|
| 6.083 | 0.43 | 0.40 | 0.154 |

**4- Conclusion**

In this article we presented the formulation method of a density functional theory MWDA for linear polymers. Some rational approximations have been used to connect MWDA method to linear polymer chain model.

For example, regarding the fact that chain polymers only crystallized when the Cohen's effect size be long enough in comparison with the range of nuclear-solid interactions between monomers so we can use the standard model of hard spheres for all of them. The connection plays a rather small role in the crystallization of linear polymer chain model and it summarized in a direct correlation function independent of temperature. Increasing the chain length increases the chance of knotting, and crystallization will be more simple, as the result the correlation function used in this research calculated just for polyethylene with N=6429. After running computer programs and doing calculations based on MWDA, which has been explained, the results have been compared with the results of famous method P-RISM and experimental results (Figs (3), (4)). As you can see from comparing in Tab. (2), our results are closer to experimental quantities both in $\rho_S$ and $\rho_l$ quantities and are more exact in comparison with the P-RISM method. The liquid phase density in MWDA method, which is equal to its experimental quantity, is 0.01 closer to its real quantity in comparison with the P-RISM method. Also the crystal phase density is 0.06 closer to real quantity in MWDA method.

As it shown in Tab. (1), our results for the length of monomers in the curve fitting method used in this research, is completely in agreement with other methods and experimental results.

Also quantities such as Lindemann criterion and chemical potential and solid and liquid phase packing fraction have been calculated in Tab. (4). Regarding the fact that they calculated based on $\rho_s, \rho_l$, we can insist on their accuracy.

According to these results, it can be explored that the method of density functional theory based on MWDA, which is used in this research, has the capability to correct the analytical results related to liquid-solid phase transition polymer linear chain model, remarkably.